\documentclass[final,3p,times]{elsarticle}

\usepackage{amssymb}

\journal{Applied Numerical Mathematics}

\newcommand{\be}{\begin{equation}}
\newcommand{\ee}{\end{equation}}
\newcommand{\bea}{\begin{eqnarray}}
\newcommand{\eea}{\end{eqnarray}}

\newcommand{\nn}{\nonumber}

\newcommand{\cO}{{\cal O}}
\newcommand{\Frac}[2]{\frac{\displaystyle #1}{\displaystyle #2}}
\newcommand{\Int}{\displaystyle{\int}}

\newcommand{\ket}{\,\rangle}
\newcommand{\bra}{\langle \,}

\begin{document}

\begin{frontmatter}



\title{Pad\'e Theory and Phenomenology of Resonance Poles}


\author{Juan Jos\'e Sanz-Cillero~\footnote{
Proceedings of the {\it Conference on
Approximation and Extrapolation of Convergent and Divergent Sequences
and Series},  CIRM Luminy, Marseille (France),
September 28th - October 2nd 2009. I would like to thank the organizers of
the conference their invitation
and the nice scientific and interdisciplinary environment.
This work is supported in part
by  CICYT-FEDER-FPA2008-01430,  SGR2009-894,
the Spanish Consolider-Ingenio 2010 Program CPAN
(CSD2007-00042), the Juan de la Cierva Program and the EU Contract
No. MRTN-CT-2006-035482, "FLAVIAnet". }   }
\address{Grup de F\'\i sica Te\`orica and IFAE,
Universitat Aut\`onoma de Barcelona,
E-08193 Bellaterra (Barcelona), Spain}


\begin{abstract}
The use of Pad\'e approximants for the description
of QCD matrix elements is discussed in this talk.
We will see how they prove to be an extremely useful tool,
specially in the case of resonant amplitudes.
It will allow the inclusion  of high-energy Euclidian data to improve the
determination of low-energy properties, such as the quadratic vector radius.
This does not mean that the rational approximations
can be arbitrarily employed for the extraction of any desired hadronic parameter.
A discussion about the validity, limitations and possible issues of the
Pad\'e analysis is carried on along the paper. Finally,
based on the de Montessus de Ballore's theorem,
a theoretically safe new procedure
for the extraction of the pole mass and width
of resonances is proposed here and
illustrated with the example of the $\rho(770)$.
\end{abstract}


\end{frontmatter}



\section{Introduction}

Quantum Chromodynamics (QCD) has been proven
to be the right theory to describe
the strong dynamics interactions.
However, in the non-perturbative regime one finds resonant structures
in the cross section which can be related with the presence of short-lived
intermediate hadronic states, usually referred as resonances.  From a mathematical
point of view,  these states appeared as complex poles of the amplitude
in the transferred energy  at higher complex Riemann sheets, other than the physical one.
Typically, the amplitude grows abruptly when the energy gets close to
the resonance pole, remaining
nevertheless  finite all the way
as the  singularity is located off the real axis, at
a complex value of the energy.

One of the clearest examples
of this kind of resonant amplitudes is the $\pi\pi$--vector
form-factor (VFF). There, the spectral function is dominated by a very pronounced
$\rho(770)$ meson peak and essentially no other big effect
is observed.
Through this qualitative knowledge of the spectral function in the
Minkowsky region ($q^2>0$)
and the experimental data from the Euclidean region ($q^2<0$)
one is able to  extract properties of the VFF at the
origin~\cite{Peris-VFF}.
Its first and second derivatives were determined at $q^2=0$
(respectively related to the quadratic  vector
radius $\bra r^2\ket_V^\pi$
and the curvature $c_V^\pi$) by means of Pad\'e approximants
(PA) centered at the origin~\cite{Peris-VFF}.   However, this precise
procedure  does not allow us to make predictions for
properties of the amplitude above threshold, such as the $\rho$--meson
pole position.

Indeed,  the potential danger of using rational
approximants for the extraction of resonance pole positions
is shown with the help of a model.   Under some limits the PAs
result  equivalent to some unitarization
procedures, such as the inverse amplitude method (IAM)~\cite{IAM,IAMb},
reason why these unitarizations have been sometimes loosely called ``Pad\'es''.
We performed a perturbative computation in the linear sigma model  (LSM)
and found that the PA around the origin led to
improper determinations of the meson mass and width~\cite{IAM-critic}.
However, we will show here that by an adequate reinterpetation of the PA
it is possible to converge (though slowly) to the actual low-energy constants
(LEC) of the model. Nonetheless,
the perturbative calculation  only makes sense in the LSM
when the $\sigma-\pi\pi$ interaction is weak.  It has been argued
that the  unitarization procedures such as IAM  are expected to work
for cases of strongly interacting mesons,
such as the physical $\sigma(600)$,  and it is not intended
for weakly interacting theories~\cite{priv-com}.
This issue is still unclear and will require of further clarifications.

Still, the Pad\'e approximants allow us
to produce a model independent determination of the resonance poles
if they are adequately employed.  To do this in a theoretically safe way,
we need to center our Pad\'e above the branch-cut singularity
(beyond the first production threshold), not bellow (at the origin).
This modification also makes possible
the direct use of Minkowsky data, being now the Euclidean ones discarded.

The importance of  these hadronic pole parameters
is that usually  one relates an  observable with
the  corresponding renormalized couplings but in the resonance
case in QCD there is still plenty of debate about which
is the right lagrangian formulation.
Alternatively, the resonance pole positions in the complex energy plane
are universal for all the processes with those same quantum numbers.
They do not depend on a particular lagrangian realization.
Nonetheless, in many cases extracting these hadronic properties brings
along much model dependence   as it is not
clear how to extrapolate  from the data on the real energy axis
into the complex plane.
This is highly non-trivial as one can see, for instance,
observing
the broad spreading of predictions for the $\sigma(600)$ meson
pole  ($I=J=0$ channel in $\pi\pi$--scattering)~\cite{PDG}.


The quadratic radius and curvature of the $\pi\pi$--VFF
will be extracted in Sec. 2 with the help of Pad\'es centered at
$q^2=0$.  A discussion on Pad\'e unitarizations is provided
in Sec. 3. Finally, a new kind of PA is proposed in Sec.~4
for the study of resonant amplitudes, centered at
energies $q^2$ over the first production threshold.

\section{Pad\'e approximants and the space-like VFF}

Our goal in this section is the description
of the $\pi\pi$--VFF   $F(Q^2)$ in the space-like  region:
\begin{equation}
\label{def}
\langle \pi^{+}(p')| \
\frac{2}{3}\ \overline{u}\gamma^{\mu}u-\frac{1}{3}\ \overline{d}\gamma^{\mu} d-
\frac{1}{3}\ \overline{s}\gamma^{\mu} s\ | \pi^{+}(p)\rangle= (p+p')^{\mu} \ F(Q^2)\ ,
\end{equation}
where $Q^2=-q^2=-(p'-p)^2$, such that $Q^2>0$ corresponds to space-like data.
Since  the spectral function for the corresponding dispersive integral for $F(Q^2)$
starts at twice the pion mass,
the form factor can be approximated by a Taylor expansion
in powers of the momentum for $|Q^2|< (2 m_\pi)^2$.

We want to construct an approximation that can be systematically improved upon.
However, it will not be our aim  to extract time-like properties from this analysis, like,
for instance, vector meson mass predictions.   It is neither our intention to
describe the amplitude on the physical absorptive cut, above
the $\pi\pi$ threshold.
Finally, it is convenient to remark that the results~\cite{Peris-VFF}
presented here do not refer to any large--$N_C$ approximation
but to the physical $N_C=3$ quantities.

\subsection{The method: Pad\'e approximants}

Consider an analytical function $F(z)$  at a  point, e.g.,  $z=0$.
A Pad\'e approximant  $P^N_M(z)=Q_N(t)/R_M(z)$
is defined by the ratio of two polynomials of degrees $N$ and $M$
which agree with the original function $F(t)$ up to the derivative
of order $N+M$ at $z=0$:
\begin{equation}
P^N_M(z)\,\,-\,\, F(z)\,\, =\,\,\cO(z^{N+M+1})\, .
\end{equation}

One may wonder what is new here with respect to a Taylor series of the form
$F(z)= a_0+a_1 z+a_2 z^2+...$
The difference is that the
polynomials are unable to go beyond the singular points. They are not able to
describe them, setting those singularities
the maximum size of the convergence disk (centered at the analytical
point $z=0$).
On the other hand, the poles of the PA tend to mimic
the singular structure of the original function $F(z)$.
For instance, if one studies the $P^N_N(z)$  approximant of $\ln{(1+z)}$
all the poles are generated at the branch cut $-\infty<z\leq -1$,
which gets  more and more densely populated as $N\to\infty$.

Thus, in many cases, the PAs are found to work far beyond the analytical
disk of convergence of the Taylor series. This allowed us to use space-like data to
improve our description of the VFF at the analytical point
$Q^2=0$.   However, in general, the Pad\'es
converge over a compact region and, although this region
may get larger and larger as one increases the order of the Pad\'e,
we did not use in this analysis information from the VFF at $Q^2=\infty$.

It is convenient to remark that in the case of non-meromorphic functions,
such as the physical amplitudes with logarithmic branch-cuts,
the poles of the PA centered at the origin
do not correspond to  resonances, but rather to  bumps and
other structures in the spectral function over the
logarithmic branch cut, which the PA tends to mimic.

In our phenomenological analysis of the VFF,
we will use as inputs all the available data in the Euclidean region, which
range from $Q^2=0.01$~GeV$^2$ up to 10~GeV$^2$.  We will also make use of
the qualitative knowledge we have on $\pi\pi$--VFF spectral function
$\rho(s)$, essentially provided by the $\rho(770)$ peak. This
suggests the use of the $P^L_1$ sequence for the description of this
particular observable.  The final aim of the analysis will be
the extraction of the first and second derivatives at $q^2=0$,
i.e., the quadratic vector radius $\bra r^2\ket_V^\pi$ and
the curvature $c_V^\pi$. From this perspective, the vector meson dominance expression
$F(Q^2)=(1+Q^2/M^2)^{-1}$ is just a $P^0_1$ Pad\'e, the first term of
a $P^L_1$ sequence.

\subsection{Theoretical uncertainties:
playing with a phenomenological model}\label{sec:model}

In order to illustrate the usefulness of the PAs
as fitting functions,
we will first use a phenomenological model as a theoretical
laboratory to check our method. The model will
also give us an idea about the size of possible systematic uncertainties.

We will consider a VFF phase-shift with the right threshold behavior
and   roughly  the physical values of the rho mass and width.
The form-factor
is recovered through a once-subtracted Omn\'es relation,
\begin{equation}\label{model}
    F(Q^2)=\exp\left \{-\frac{Q^2}{\pi} \int_{4 \hat{m}_{\pi}^{2}}^{\infty}\ dt\ \frac{\delta(t)}{t (t+Q^2)}\right\}\ ,
\end{equation}
where $\delta(t)$ plays the role of the vector form factor
phase-shift~\cite{Pich-VFF,Cillero-VFF,Portoles} and  is given by
\begin{equation}\label{model2}
    \delta(t)=\tan^{-1}\left[\frac{\hat{M}_{\rho}
    \hat{\Gamma}_{\rho}(t)}{\hat{M}_{\rho}^2-t} \right]\ ,
\end{equation}
with the $t$-dependent width given by
\begin{equation}\label{width}
    \hat{\Gamma}_{\rho}(t)= \Gamma_{0}\ \left( \frac{t}{\hat{M}_{\rho}^2} \right)\ \frac{\sigma^3(t)}{\sigma^3(\hat{M}_{\rho}^2)}\ \theta\left( t- 4 \hat{m}_{\pi}^{2} \right)\ ,
\end{equation}
and $\sigma(t)=\sqrt{1-4 \hat{m}_{\pi}^{2}/t}$.
The input parameters  are chosen to be close to their physical values:
$ \Gamma_{0} = 0.15\ \mathrm{GeV}$,
 $\hat{M_{\rho}}^2= 0.6\ \mathrm{GeV}^2$,
 $4 \hat{m}_{\pi}^{2}= 0.1 \ \mathrm{GeV}^2$.
%
%
This model is actually  quite realistic and
it has been used  for the extraction of  the
physical mass and width of the $\rho(770)$  meson
from time-like experimental data~\cite{Pich-VFF,Cillero-VFF,Portoles}.

We generate now an emulation of the experimental data from our theoretical
model. In order to recreate the situation of the experimental
data~\cite{Amendolia}-\cite{Dally} with the model, we have generated fifty ``data'' points
in the region $0.01\leq Q^2\leq 0.25$, thirty data points in the interval $0.25\leq Q^2 \leq 3$, and seven
points for $3\leq Q^2\leq 10$ (all these momenta in units of GeV$^2$). These points are taken with vanishing
error bars since  our purpose here is  to estimate the systematic error derived purely from our approximate
description of the form factor.

These generated data is then fitted through $P^L_1$ Pad\'e approximants,
\begin{equation}
P_1^L(Q^2) \,\, \,= \,\,\,
1\, +\,  \sum_{k=0}^{L-1}a_k (-Q^2)^{k} \,\,
+ \, (-Q^2)^{L} \,  \frac{\ a_L}{1+\Frac{a_{L+1} }{a_L}  \, Q^2}\  ,
\label{PL1}
\end{equation}

where the vector current conservation condition $P^L_1(0)=1$,
i.e., $a_0=1$, has been imposed.
At low energies this produces for the Taylor
coefficients $a_j$ the prediction
\begin{equation}\label{expmodel}
    F(Q^2)
    \, =\, 1 \, - \,  a_1\ Q^2 \, + \,  a_2\ Q^4 \,  - \ a_3\ Q^6 + ...
\end{equation}

This leads to a series of  predictions for the low-energy parameters,
which are compared to their (known) exact values in Table~\ref{table1}.
The last PA we have fitted to these data is $P^6_1$. Notice
that the pole position of the Pad\'{e}s differs from the
mass parameter of the model $\hat{M}_\rho^2$
--and this from the pole mass--.
This example makes explicit how one can get a rather precise
value for the Taylor coefficients in Eq.~(\ref{expmodel}) without an accurate knowledge
of the spectral function (i.e., of the time-like region).

\begin{table}[t]
\centering
\begin{tabular}{|c|c|c|c|c|c|c|c|c|}
  \hline
   & $P^{0}_{1}$ &  $P^{1}_{1}$ &  $P^{2}_{1}$ &  $P^{3}_{1}$&$ P^{4}_{1}$  & $P^{5}_{1}$  &$P^{6}_{1}$ & $F(Q^2)$(exact)\\ \hline
  $a_1$(GeV$^{-2}$) & 1.549 & 1.615 & 1.639 & 1.651 & 1.660&1.665 & 1.670 & 1.685 \\
  $a_2$ (GeV$^{-4}$)& 2.399 & 2.679& 2.809 & 2.892 & 2.967&3.020 & 3.074& 3.331\\
  $a_3$(GeV$^{-6}$)& 3.717 & 4.444 & 4.823 & 5.097 & 5.368&5.579 & 5.817& 7.898\\
  \hline
  \hline
  $s_p$(GeV$^{2}$) &$0.646$&$0.603$&$0.582$&$0.567$&$0.552$&$0.540$
  &$0.526$&  $\hat{M}_\rho^2=0.6$\\
  \hline
\end{tabular}
\caption{{\small Results of the various fits to the form factor $F(Q^2)$ in the model, Eq. (\ref{model}).
The exact values for the coefficients $a_i$ in Eq. (\ref{expmodel}) are given on the last column. The last
row shows the predictions for the corresponding pole for each Pad\'{e}
($s_p$), to be compared to the mass parameter
$\hat{M}_{\rho}^{2}=0.6\ $GeV$^2$ in the model.}} \label{table1}
\end{table}

Based on the previous results, we will take the values in
Table \ref{table1} as a rough estimate of the systematic
uncertainties when fitting to the experimental data in the
following sections. Since, as we will see, the best fit to the
experimental data comes from the  Pad\'{e} $P^4_1$, we will take the
error in Table \ref{table1} from this Pad\'{e} as a reasonable estimate
and, respectively,
add to the final error  an extra systematic uncertainty
of $1.5\%$ and $10\%$  for $a_1$ and $a_2$.

\subsection{Experimental pion vector  form factor}
\label{sec:PAL1}

\begin{figure}[!t]
  \center
  \includegraphics[width=5cm]{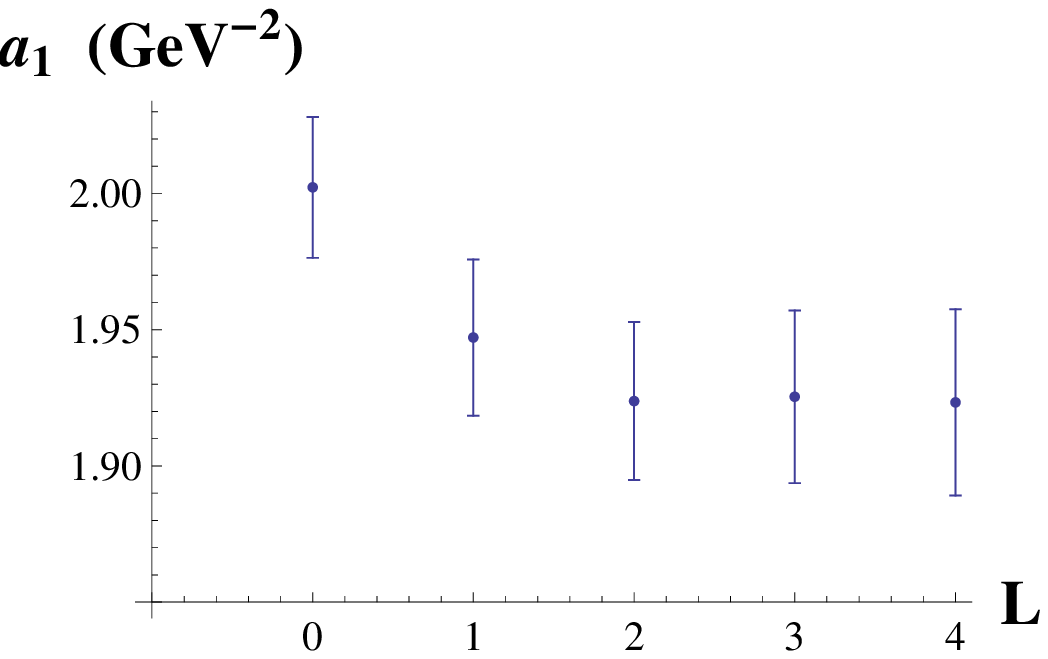}
  \hspace*{1.cm}
  \includegraphics[width=5cm]{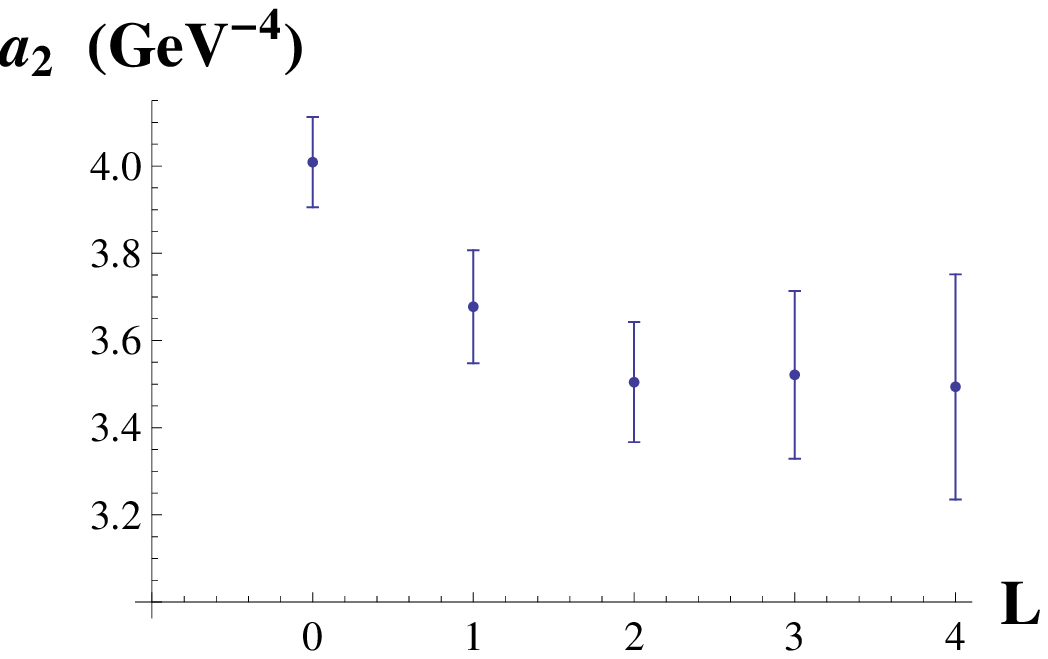}
  \caption{{\small
  Value of the $a_1$ and $a_2$ Taylor coefficients for the
  $P^L_1$  sequence of Pad\'e Approximants  obtained from
  experimental data fits~\cite{Amendolia}--\cite{Dally}.  }}
  \label{fig:a1PL1}
\end{figure}

The prominent role of the
rho meson contribution motivates
the use of the $P^{L}_{1}$ Pad\'e sequence as the central tool
for the study of this amplitude, later complemented by other types
of Pad\'es.
The fit of $P^L_1$ to the space-like data points
in Refs.~\cite{Amendolia}--\cite{Dally}
determines  the coefficients  $a_{k}$ that best interpolate them.
Fig.~\ref{fig:a1PL1} shows the evolution of the fit results for the Taylor
coefficients $a_1$ and $a_2$ for the $P^L_1$ PA  from $L=0$ up to $L=4$.
As one can see, after a few Pad\'es
these coefficients become stable.
 For the  data in Refs. ~\cite{Amendolia}-\cite{Dally}, this happened
 at $L=4$. The $P^4_1$ Pad\'e Approximant  provides
 our best fit and, upon expansion around $Q^2=0$,  this yields
\begin{equation}
a_1\, =\, 1.92 \pm 0.03\,\,\mbox{GeV}^{-2} \, ,  \qquad\qquad a_2\, =\, 3.49 \pm 0.26\,\,\mbox{GeV}^{-4} \,
;
\end{equation}
with a $\chi^2/\mathrm{dof}=117/90$~\cite{Peris-VFF}.

\begin{figure}
  \center
  \includegraphics[width=5cm]{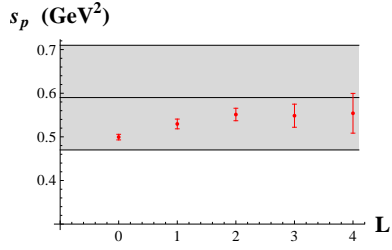}
  \caption{{\small Position $s_p$ of the pole for the different $P^L_1$.
  The range with the physical values
  $M_\rho^2\pm M_\rho\Gamma_\rho$ is shown (gray band)
  for comparison. }}\label{fig:spPL1}
\end{figure}

Eq.~(\ref{PL1}) shows that the pole of each $P^L_1$ PA is determined by the ratio $s_p=a_L/a_{L+1}$.  This
ratio is shown in  Fig.~\ref{fig:spPL1}, together with a gray band
given by $M_\rho^2 \pm M_\rho\Gamma_\rho$ for comparison. From this figure
 one can see that the position of the pole of the PA  is close to the
physical mass $M_\rho^2$~\cite{PDG}, although it cannot be
identified with it,
as we already saw in the model of the previous subsection.
The $P^L_1$ pole $s_p$ is always real and lies at the ``bump''
of the spectral faction, in the range $M_p^2\pm M_p\Gamma_p$.
The Pad\'e tends to reproduce the $\rho$ peak line-shape but, obviously,
no complex resonance pole can be recovered from a $P^L_1$ Pad\'e
at higher Riemann sheets.

\begin{figure}
  \center
  \includegraphics[width=5cm]{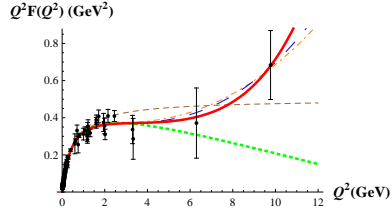}
  \caption{{\small The sequence of $P^L_1$ PAs is compared to the available space-like
  data~\cite{Amendolia}-\cite{Dally}:
  $P^0_1$ (brown dashed), $P^1_1$ (green thick-dashed),
  $P^2_1$ (orange dot-dashed), $P^3_1$ (blue long-dashed),
  $P^4_1$ (red solid).}}
  \label{fig:VFF}
\end{figure}

As one can see in Fig.~\ref{fig:VFF}, the sequence $P^L_1$
converges to the physical form-factor  in the data region but,
eventually,  it diverges   like $(Q^2)^{L-1}$.
These PAs only converge on a compact region of the complex plane
and are unable to recover the $1/Q^2$ asymptotic behaviour prescribed
by QCD at short distances~\cite{brodsky-lepage}.
Nonetheless, the important fact is that the Pad\'es allow the use of
data not only by the origin but even from energies as large as
$Q^2\sim 10$~GeV$^2$, something that a normal Taylor expansion at the origin
does not permit.

\begin{table}[t!]
\centering
\begin{tabular}{|c|c|c|c|c|c|c|c|c|}
  \hline
   &  $PA^{L}_{2}$ &  $PT^L_1\,\, (\rho)$
   &  $PT^L_2\,\,(\rho,\rho')$ &  $PT^L_2\,\,(\rho,\rho'')$
   &   $PT^L_3\,\,(\rho,\rho',\rho'')$
   &  $ PP^L_{1,1}\,\, (\rho)$
   \\
  & & &  & & &
    \\   \hline
  $a_1$(GeV$^{-2}$)
  & $1.924\pm 0.029$ & $1.90\pm 0.03$ & $1.902\pm 0.024$ &
  $1.899\pm 0.023$ & $1.904\pm 0.023$ & $1.902\pm 0.029$
  \\
  $a_2$ (GeV$^{-4}$)
  & $3.50\pm 0.14$   & $3.28\pm 0.09$  &
  $3.29\pm 0.07$  &  $3.27\pm 0.06$    &   $3.29\pm 0.09$  &
  $3.28\pm 0.09$ \\
  \hline
\end{tabular}
\caption{{\small Different results for two-pole Pad\'e approximants,
Pad\'e-types and partial Pad\'es, where we used the $\rho$, $\rho'$, and
$\rho''$ masses.}}
\label{tab.Pades}
\end{table}

\begin{table}[b!]
\begin{center}
\begin{tabular}{|c|c|c|c|c|c|c|c|c|}
  \hline
   & $\langle r^2\rangle_V^\pi$   (fm$^2$)  &  $a_2$  (GeV$^{-4}$)     \\ \hline \hline
  This work &   $0.445\pm 0.002_{\mathrm{stat}}\pm 0.007_{\mathrm{syst}}$
  &  $3.30\pm 0.03_{\mathrm{stat}}\pm 0.33_{\mathrm{syst}}$    \\
   PDG~\cite{PDG} & $0.452\pm 0.011$  & ... \\
  CGL~\cite{Colangelo,ColangeloB}& $ 0.435\pm 0.005$  & ...   \\
  TY~\cite{Yndurain}  & $ 0.432\pm 0.001 $  & $ 3.84\pm 0.02$ \\
  BCT~\cite{op6-VFF} & $0.437\pm 0.016$  & $3.85\pm 0.60$ \\
  PP~\cite{Portoles} & $0.430\pm 0.012$  & $3.79\pm 0.04$ \\
   Lattice~\cite{lattice} & $0.418\pm 0.031$  & ... \\
  \hline
\end{tabular}
\caption{{\small Our results for the quadratic vector radius
$\langle r^2\rangle_V^\pi$ and second derivative $a_2$ are
compared to other
determinations~\cite{PDG,Portoles,Colangelo,ColangeloB,Yndurain,op6-VFF,lattice}. Our first error is
statistical. The second one is systematic, based on a
previous analysis of a VFF model.}}
\label{table2}
\end{center}
\end{table}

This $P^L_1$ PA analysis was complemented with the results from
other types of Pad\'es: two-pole Pad\'e approximants $PA^L_2$;
one, two and three-pole Pad\'e-types (PT)
with the poles fixed beforehand; two-pole partial Pad\'es $PP^L_{1,1}$,
with one pole fixed beforehand and the other determined by the low-energy
coefficients $a_k$.  The different results are gathered in
Table.~\ref{tab.Pades}.

Combining all the previous rational approximants results
in the  average given by~\cite{Peris-VFF}
\begin{equation}
a_1\, =\, 1.907\pm 0.010_{\mathrm{stat}} \pm 0.03_{\mathrm{syst}}\,\,\mbox{GeV}^{-2} \  ,  \
a_2\, =\,  3.30 \pm  0.03_{\mathrm{stat}} \pm 0.33_{\mathrm{syst}}\,\,\mbox{GeV}^{-4} \, .
\end{equation}
The first error comes from combining the results from
the different fits by means of a weighted average. On
top of that, we have added what we believe to be a conservative
estimate of the theoretical (i.e.
systematic) error based on the analysis of the VFF model
in the previous subsection. We expect the latter to
give an estimate for the systematic uncertainty due
to the approximation of the physical form factor with
rational functions. For comparison with previous analyses,
we also provide in Table~\ref{table2} the value
of the quadratic vector radius,
which is given by $\langle r^2 \rangle_V^\pi \, =\, 6 \, a_1$ .

In summary,  we  used rational approximants
as a tool for fitting the pion vector form factor
in the Euclidian range.
Since  these approximants are capable of
describing the region of large momentum, we think they are
better suited than polynomials for the description of the
currently available  space-like data.
As our results in Table~2 show, the errors achieved with
these approximants  are competitive with previous analyses
existing in the literature, based on more elaborated techniques.

\section{A critical look on Pad\'e unitarizations}

Exact unitarity is an important additional piece of information
that eventually needs to be included in the description of the
scattering processes. A commonly employed unitarization
procedure is the inverse amplitude method
(IAM)~\cite{IAM,IAMb,IAM-P12-critics,IAM-poles,IAM-op6}.
Although it can be formulated in a more elaborated way through dispersion
relations, the method relies in the unitarity relation in the elastic
region, which in the massless case has the simplified form,
\begin{equation}
\mbox{Im}t(s)\,=\, |t(s)|^2\qquad\longrightarrow\qquad
\mbox{Im[}t(s)^{-1}\mbox{]}\,=\, -1\,.
\end{equation}
This fixes completely the imaginary part of the
inverse partial-wave amplitude $t(s)^{-1}$ on the
elastic part of the right-hand cut ($s>0$).  All that remains
in $t(s)^{-1}=\mbox{Re[}t(s)^{-1}\mbox{]}+\mbox{Im[}t(s)^{-1}\mbox{]}$
is to determine the real part of $t(s)^{-1}$, which is fixed in the IAM
through a low-energy matching to $\chi$PT:
\begin{equation}
t(s)_{\chi PT}\,=\, t_{(2)}\, +\,t_{(4)}\, +\, t_{(6)}\, +\,\, ...
\end{equation}
where the $t_{(k)}$ are the contributions corresponding to $\cO(p^k)$
in $\chi$PT.
Hence, depending on the order of the matching one obtains a sequence
of unitarized amplitudes~\cite{IAM,IAMb,IAM-P12-critics,IAM-poles,IAM-op6}:
\begin{equation}
\cO(p^4)\longrightarrow\quad
t_{IAM}\,=\, \Frac{t_{(2)}}{1\,-\, t_{(4)}/t_{(2)}} \, ,
\qquad\qquad
\cO(p^6)\longrightarrow\quad
t_{IAM}\,=\, \Frac{t_{(2)}}{1\,-\, t_{(4)}/t_{(2)}\, -\, t_{(6)}/t_{(2)}
\, +\, (t_{(4)}/t_{(2)})^2 } \, ,\quad ...
\label{eq.IAM-scat}
\end{equation}
Identical results are obtained if one recovers the partial wave amplitude
through a dispersion relation and matches $\chi$PT on the left-hand cut.

In the tree-level limit,
the IAM expressions~(\ref{eq.IAM-scat}) become a series of
Pad\'e approximants of the form $P^1_1$, $P^1_2$, etc, reason why
these  unitarizations are sometimes called
``Pad\'es''.    In any case, it is in this limit that one can use all the
powerful  technology of the mathematical theory of Pad\'e approximants.
However, it has been argued that the IAM should be only applicable
for the description of broad resonances, such as the sigma
meson, and not narrow states~\cite{priv-com}.

The IAM has been found to described the data reasonably
well even, in some cases, up to energies as high
as $\sqrt{s}\sim 1$~GeV~\cite{IAM,IAMb}.
It has been able to generate
all the  resonances below 1~$GeV$: $\rho$, $K^*$, $\sigma$
(or $f_0(600)$)$, \kappa$, $a_0(980)$ and
$f_0(980)$~\cite{IAM,IAMb,IAM-poles,IAM-op6},
and the expected $N_C$ behaviour of
the $\rho$ and $K^*$ poles as $q\bar{q}$ states
has been recovered~\cite{IAM-poles}.

However, one may wonder what information is lost when Re[$t(s)^{-1}$]
is fixed at low energies with $\chi$PT. Likewise,
in general the unitarized amplitudes violate crossing, as only
it only resums a particular set of diagrams in the $s$--channel, not in
the crossed ones.
Nevertheless,
even though the IAM determinations have produces good numerical results,
this is not supported by the theory of Pad\'e approximants.
Thus, we consider that the
theoretical reason why the IAM has been so successful at the
phenomenological level still needs and deserves further clarifications.

\subsection{A counterexample: The $\sigma$ in the Linear Sigma Model}

The application of the IAM to the linear sigma model (LSM)
was proposed in Ref.~\cite{IAM-critic} as a counterexample,
to show how the prediction from unitarizing the low-energy LSM
led to very different pole mass and width than those
actually in the LSM.

At tree-level,  the sigma mass is
found to be $M_\sigma^2=2 \mu^2$, and the width is zero. At
next-to-leading order, the sigma pole gets shifted due to the
quartic potential, i.e., $M_\sigma^2= 2\mu^2 + \cO(g)$, and the
width becomes different from zero. For simplicity,
the massless pion limit is assumed.

In order to determine the scalar meson mass and width up to $\cO(g)$,
we compute the one-loop sigma correlator~\cite{oneloop},
\begin{equation}
i\Delta(s)^{-1}\, =\, s\, -\, M_\sigma^{2} \, \left[1 + \Frac{3 g}{16\pi^2}\, \left( -\Frac{13}{3} +  \ln\Frac{-s}{M_\sigma^2} + 3 \rho(s) \ln{\left(
\Frac{\rho(s)+1}{\rho(s)-1}\right)}  \right) + \cO(g^2) \right]\, ,
\end{equation}
where $\rho(s)\equiv \sqrt{1- 4 M_\sigma^2/s}\ $  and the term  $-13/3$
is determined by the renormalization scheme chosen by Ref.~\cite{oneloop},
which sets the relation $2 g F^2=M_\sigma^2$ at the one-loop order, with $F$ the
pion decay constant and $M_\sigma$ the renormalized mass parameter.
Now it is
possible to extract the pole $s_p$ of the propagator up to the considered order in perturbation theory. If one approaches the branch cut from the
upper part of the complex $s$--plane, the pole in the second
Riemann sheet is located at  $s_p=(M_p-i \Gamma_p/2)^2$,
with the pole mass and width,
\begin{eqnarray}
\left(\Frac{M_p^2}{M_\sigma^{  2} }\right)_{_{LSM}}
&=& 1 \, +   \Frac{3 g}{16\pi^2}\, \left( -\Frac{13}{3} +  \pi\sqrt{3}  \right)+ \cO(g^2)\, \nonumber \\[2mm]
\left(\Frac{M_p \,\Gamma_p}{M_\sigma^{  2}}\right)_{_{LSM}} &=&  \Frac{3 g }{16\pi} \,+\, \cO(g^2) \, .
\label{eq.LSM-pole}
\end{eqnarray}

The $\pi\pi$--scattering is determined by the
$\pi^+\pi^- \to \pi^0\pi^0$ amplitude $A(s,t,u)$.
This defines the isospin amplitudes
\begin{eqnarray}
T(s,t,u)^{\rm I=0} &=&
3 A(s,t,u) + A(t,s,u) +A(u,t,s) \, ,
\nn \\
T(s,t,u)^{\rm I=1} &=&
A(t,s,u) - A(u,t,s)\, ,
\nn \\
T(s,t,u)^{\rm I=2} &=&
A(t,s,u)+A(u,t,s) \, ,
\end{eqnarray}
and the partial wave projection provided by
\begin{equation}
t^I_J(s) \,\, =\,\, \Frac{1}{64\pi} \Int_{-1}^1 d\cos{\theta}
\, \, P_J(\cos{\theta})\,\,  T(s,t,u)^{\rm I}\, ,
\label{eq.PW}
\end{equation}
where $\theta$ is the scattering angle in the $\pi\pi$ center-of-mass rest frame.

We now consider the LSM at low energies, which reproduces
the structure prescribed by  $\chi$PT.
Hence, for the first partial waves $t^I_J(s)$, with $IJ=00,11,20$,
it produces the  $\cO(p^2)$ amplitudes,
\begin{eqnarray}
t_0^0(s)_{(2)}  =  \Frac{s}{16\pi F^2} \, , \qquad \qquad
t_1^1(s)_{(2)}  =  \Frac{s}{96\pi F^2}\, , \qquad \qquad
t_0^2(s)_{(2)}  =  -\Frac{s}{32\pi F^2}\, ,
\label{eq.t2}
\end{eqnarray}
and at $\cO(p^4)$,
\begin{eqnarray}
t_0^0(s)_{(4)}\, &=& \, t_0^0(s)_{(2)} \, \, \times\,\, \Frac{11 s}{6 M_\sigma^{  2}} \left[  1  - \Frac{g}{264\pi^2} \left(
18\ln\Frac{-s}{M_\sigma^{  2}} + 7 \ln\Frac{s}{M_\sigma^{  2}} +\Frac{193}{3}\right) +\cO(g^2) \right] \, , \nn \\[2mm]
t_1^1(s)_{(4)} \, &=&\, t_1^1(s)_{(2)} \,\, \times\,\, \left(\Frac{-s}{M_\sigma^{  2}}\right)\,\, \left[1  + \Frac{g}{48\pi^2} \left(
\ln\Frac{-s}{M_\sigma^{  2}} - \ln\Frac{s}{M_\sigma^{ 2}} -\Frac{26}{3}\right) +\cO(g^2) \right] \, , \nn \\[2mm]
t_0^2(s)_{(4)} &=& t_0^2(s)_{(2)} \,\,\times\,\, \left( \Frac{- 2 s}{3 M_\sigma^{  2} } \right)\,\, \left[  1  - \Frac{g}{24\pi^2} \left(
\Frac{9}{4} \ln\Frac{-s}{M_\sigma^{  2}} + \Frac{11}{4} \ln\Frac{s}{M_\sigma^{  2}} +\Frac{163}{24}\right)+\cO(g^2) \right] \, .
\label{eq.t4}
\end{eqnarray}


The Inverse Amplitude Method (IAM
) provides an amplitude that is unitary not only at the perturbative
level but exactly.
At $\cO(p^4)$, one has the unitarized amplitude,
$ t_{_{\rm IAM}}\, \, =\, \,
\frac{t_{(2)}}{1\, -\, \Frac{t_{(4)}}{t_{(2)}}}$,
which has its poles $s_p$ at  $t_{(2)}(s_p)=t_{(4)}(s_p)$:
\begin{eqnarray}
\mbox{\bf IJ=00}\longrightarrow\qquad s_p&=&   \Frac{ 6}{11}  M_\sigma^{  2} \, \left[  1  +  \Frac{g}{264\pi^2}
\left( \Frac{193}{3} + 25 \ln\Frac{6}{11} - 18 i \pi \right)
+\cO(g^2) \right] \, \, ,
\\
\mbox{\bf IJ=11}\longrightarrow\qquad s_p &=&  - M_\sigma^{  2} \,\,
\left[1  +  \Frac{g}{48\pi^2}
\left(   \Frac{26}{3} + i\pi \right)
+\cO(g^2) \right] \, ,
\\
\mbox{\bf IJ=20}\longrightarrow\qquad s_p &=&    -\Frac{3}{2} M_\sigma^{  2} \,\,
\left[  1  + \Frac{g}{24\pi^2}
\left( \Frac{163}{24}+5 \ln\Frac{3}{2}+ \Frac{11 i \pi}{4}\right)+\cO(g^2) \right]
\, .
\label{eq.LSM20}
\end{eqnarray}
These are the poles that appear in the unphysical Riemann sheet
as one approaches from upper half of the first Riemann sheet.
There is also a
conjugate pole at $s_p^*$ if one approaches the real $s$--axis from below.

The first thing to be noticed is that poles appear in the $IJ=11$ and $20$
channels even for small values of $g $, contrary to what one expects in
the LSM, where no meson  with these quantum numbers exists.
Furthermore,
these ``states'' are not resonances, as they are located on the left-hand
side of the complex $s$--plane, out of the physical Riemann sheet,
and carrying a negative squared mass.

One can easily see the important
disagreement between the actual LSM value for  the
pole mass and width given in Eq.~(\ref{eq.LSM-pole})
and the posterior IAM ``prediction'',
\begin{eqnarray}
\left(\Frac{M_p^2}{M_\sigma^2}\right)_{_{IAM}}&=&
\Frac{6}{11} \, \, \left[  1  +  \Frac{g}{16\pi^2} \left( \Frac{50}{33} \ln\Frac{6}{11} +\Frac{386}{99}\right)
+\cO(g^2) \right] \, \, , \nn \\[2mm]
\left(\Frac{M_p \, \Gamma_p}{M_\sigma^2}\right)_{_{IAM}}
&=& \Frac{24}{121}\,\cdot \,  \Frac{ 3 g}{16\pi}\,\, +\, \cO(g^2)  \, ,
\end{eqnarray}
this is, $(M_p^2)_{_{IAM}}\simeq 50\%\, (M_p^2)_{_{LSM}}$
and $(M_p\Gamma_p)_{_{IAM}}\simeq 20\% \,(M_p\Gamma_p)_{_{LSM}}$.

\subsection{Higher order Padé Approximants for the LSM}
\label{padessec}

\quad In the tree-level limit (for instance, at large $N_C$),
the IAM amplitudes become Pad\'e  approximants centered at~$s=0$:
\begin{eqnarray}
 \mbox{[} \cO(p^4)\mbox{]}\quad
 \Frac{t_{(2)}}{1\,-\, t_{(4)}/t_{(2)}}
\,\,=\,\, t_{(2)}\,+\,t_{(4)}\,+\, ...
\qquad\qquad &\longrightarrow&  \qquad\qquad P^1_1(s)
\nn\\
 \mbox{[} \cO(p^6)\mbox{]}\quad
 \Frac{t_{(2)}}{1\,-\, t_{(4)}/t_{(2)}\, -\, t_{(6)}/t_{(2)}
\, +\, (t_{(4)}/t_{(2)})^2 }
\,\,=\,\, t_{(2)}\,+\,t_{(4)}\,+\,t_{(6)}\,+\, ...
\qquad\qquad  & \longrightarrow &  \qquad\qquad P^1_2(s)
\nn\\
&&\qquad\qquad ...  \nn
\end{eqnarray}

The $\pi\pi$--scattering is given in the LSM case by
\begin{eqnarray}
A(s,t,u)\,\, =
\,\, \Frac{s}{F^2}\,\Frac{M_\sigma^2}{M_\sigma^2\,-\,s}
\qquad\longrightarrow\qquad
t_0^0(s)  &=& \Frac{M_\sigma^2}{32 \pi F^2} \, \left[  - 5 + \Frac{ 3M_\sigma^2}{M_\sigma^2-s} + \Frac{ 2 M_\sigma^2}{s}
\ln\left(1+\Frac{s}{M_\sigma^2}\right)\right] \, ,
\end{eqnarray}
which at low energies become,
\begin{eqnarray}
A(s,t,u)\,\, =\,\, \Frac{s}{F^2}\,\left[ 1\, +\, \Frac{ s}{M_\sigma^2} \,
+\, \Frac{ s^2}{M_\sigma^4} \, \, +\,\, ...\right]
\qquad \longrightarrow\qquad
t_0^0(s)  &=& \Frac{s}{16 \pi F^2} \,
\left[ 1 + \Frac{11 s}{6 M_\sigma^2} +\Frac{15 s^2}{12 M_\sigma^4}\,\,
+\,\,...\right] \, .\end{eqnarray}

\begin{figure}[!t]
\begin{center}
  \includegraphics[width=5cm]{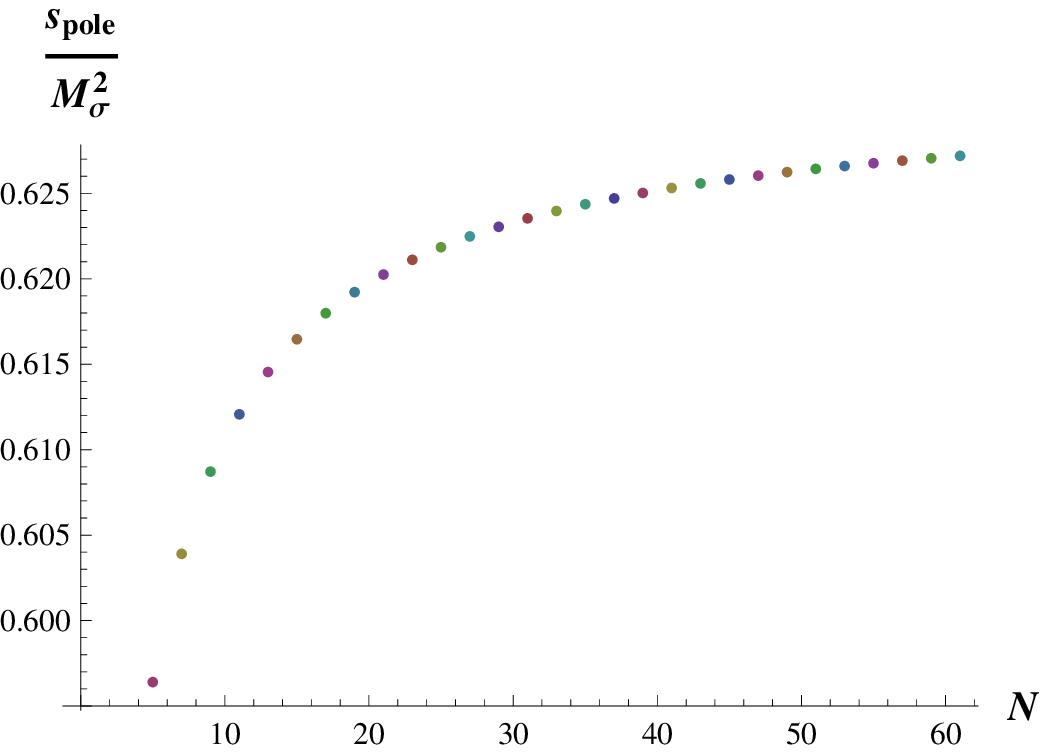}
  \hspace{2cm}
  \includegraphics[width=5cm]{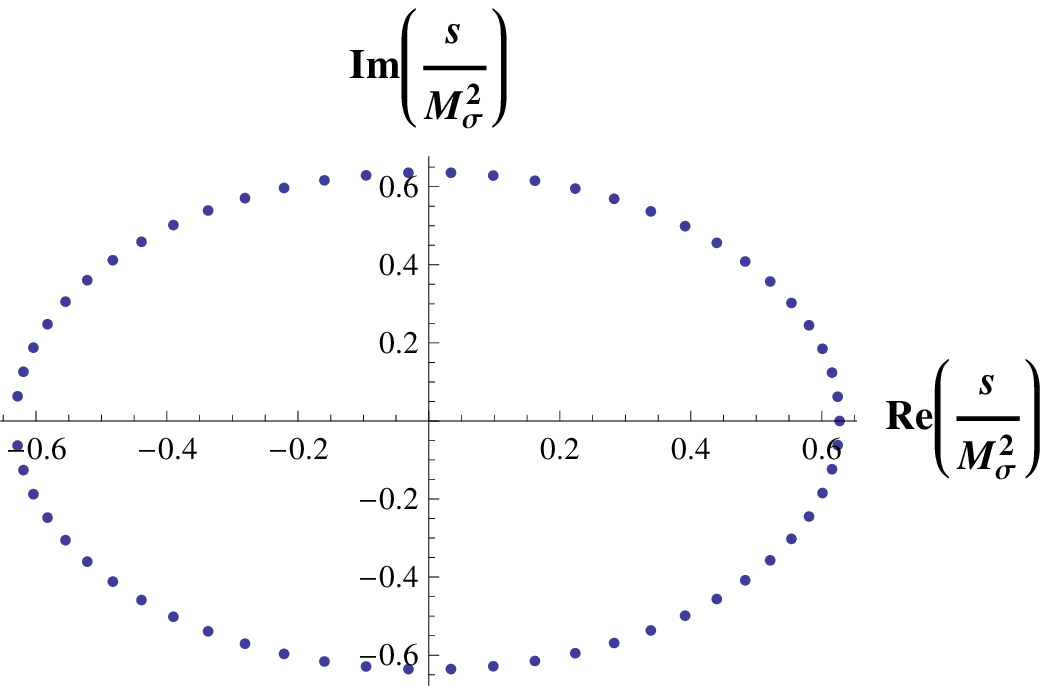}
  \caption{{\small   (a) Position of the nearest pole to $M_\sigma^2$
  for the first PAs of the form $P^1_N$ with $N$ odd
  (for even $N$ all the poles are complex).
  (b) Poles of the $P^1_{61}$ in the complex plane.}}
  \label{LSM30}
\end{center}
\end{figure}

The first Pad\'e approximant, $P^1_1$, gives for the $\sigma$ pole
the prediction $s_p=\frac{6}{11}M_\sigma^2$.  But how does the series
$P^1_1$, $P^1_2$, $P^1_3$... evolve?  One could even wonder,
for instance,  about
the behaviour of other sequences such as $P^1_1$, $P^2_2$, $P^3_3$...
We have shown in Fig.~\ref{LSM30} the $P^1_M$ sequence up to such a high order
as $P^1_{61}$. One can see the extremely slow converges, with the pole prediction
still a 30\% off for $P^1_{61}$. Furthermore,
this kind of $P^1_M$
Pad\'e approximants has only one zero (at $s=0$) and
places the $M$ poles in a circle centered at $s=0$,
producing an analytical structure completely different to
that in the actual LSM.

\begin{figure}[!t]
\begin{center}
  \includegraphics[width=5cm]{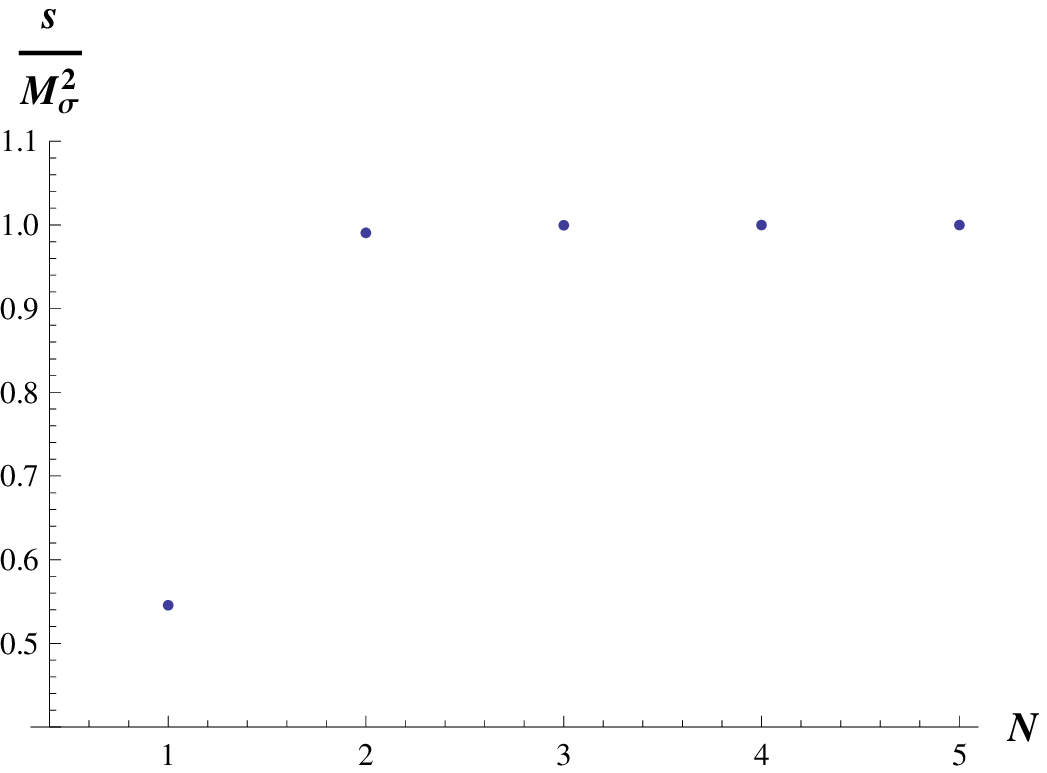}
  \hspace{1.5cm}
  \includegraphics[width=5cm]{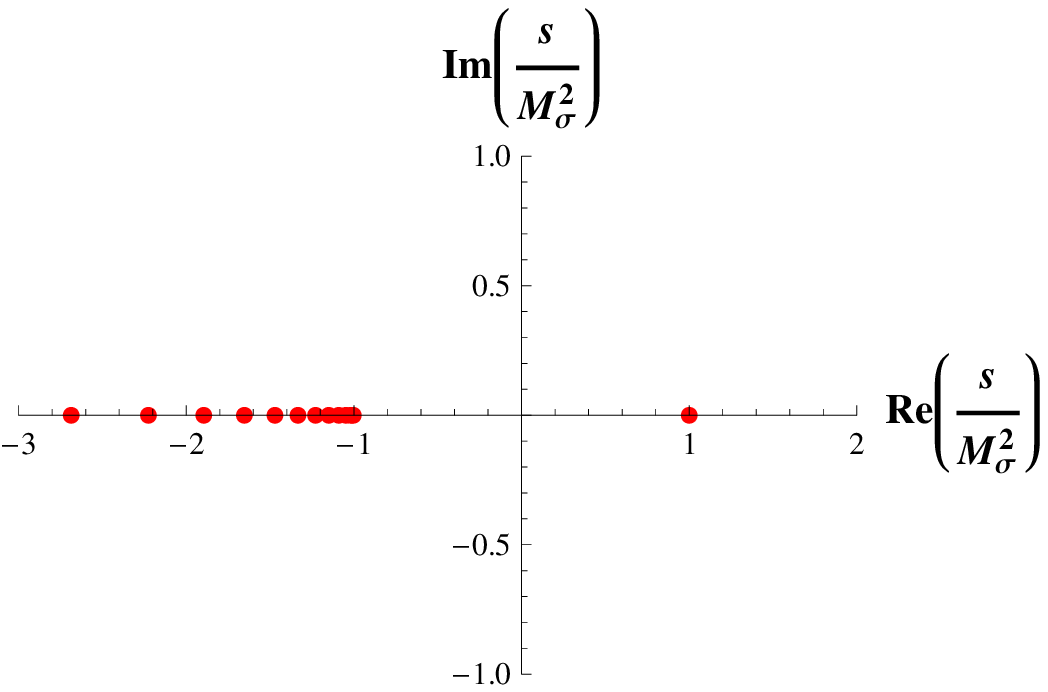}
  \caption{{\small
  (a) Location of the closest pole to $M_\sigma^2$
  for the first $P^N_N$ Pad\'e approximants.
  (b) Poles of $P^{20}_{20}$.
  }}
\label{PadeNN20}
\end{center}
\end{figure}

On the other hand, a quick convergence is found for the diagonal Pad\'e
sequence $P^N_N$:   $P_1^1$ reproduces the sigma pole a
$40\%$ off but $P^2_2$ disagrees by less than $1\%$, $P^3_3$ by less
than $0.1\%$, etc.
Likewise,  Fig.~\ref{PadeNN20}.b shows how the $P^N_N$ sequence,
besides providing the isolated pole of the sigma,
tends to reproduce the left-hand cut as $N$ increases.
The poles of $P_{20}^{20}$ are plotted there.
Although a PA is a rational function without cuts,
these are mimicked by placing poles where the cuts should lie.
The $P_{20}^{20}$ has one isolated pole near $M_{\sigma}^2$
(with an accuracy of $10^{-30}$) and nineteen poles over the
real axis at $s_p<-M_{\sigma}^2$, i.e. on the left-hand cut
of the original function. As $N$ is increased, the number of
poles lying on the branch cut increases too.
A last remarkable feature is that the $P_N^N$ approximants
obey exact unitarity, as it happened with the IAM sequence $P^1_N$.

However, phenomenologically, the IAM  has been usually employed for the
study of the experimental data in the resonance region.  These have
been used to  fit its parameters (e.g.,  the $\hat{\ell}_i$ in
Ref.~\cite{IAMb})  and to extract the resonance pole positions.
These parameters also  provide a prediction for the
chiral low-energy constants.
However, although in general  the
convergence of these parameters to the LECs is unclear,
in the case of the sequence of one-pole Pad\'es $P^N_1$ centered at
$s\sim  M_\sigma^2$,  the convergence is ensured in the
disk centered at that point and  with maximal
radius limited by the position
of the essential singularity of the left-hand cut at
$s=-M_\sigma^2$~\cite{Montessus,BakerPades}.
Hence, the predictions of the $P^N_1$ Pad\'e approximant
at the middle point $s=0$ converge to the actual $\chi$PT  low-energy
couplings: the $P^1_1$ centered at $s=M_\sigma^2$ recovers the
$\cO(p^4)$ coefficient with a 50\% error,
31\% for $P^2_1$, 17\% for $P^3_1$, etc.   From this perspective, maybe one
could explain the phenomenological success of the IAM
(which shares the first term, $P^1_1$),
although  the $P^N_1$ sequence might be more adequate
for low-energy predictions than the usual IAM pattern $P^1_N$.

\section{Model independent determination of the resonance poles}

As we have seen in previous sections, the Pad\'es have been used to extract
low-energy parameters from Euclidean data. In a missleading way,
some unitarizations procedures have been called Pad\'es, although
their justification  from the point of view of Pad\'e theory is
unclear.  Though these unitarizations have been proven more or less
successful in their predictions of the resonance poles and low-energy
coefficients, they are not supported by
the mathematical theory of Pad\'es.
Alternatively, in this section we show how it is possible to construct a
different type of Pad\'e approximants which allows us to use
Minkowskian data and to extract the resonance pole position
through  a theoretically safe procedure supported by mathematical theorems.

To illustrate the procedure, let us start by the simplest possible case.
If one has a function $F(s)$ analytical in a disk
$B_\delta (s_0)$  then the Taylor series
$S_N(s)=\sum_{k=0}^N a_k (s-s_0)^k$ converges to $F(s)$ in
$B_\delta (s_0)$ for  $N\to\infty$,  with the derivatives
$a_k=F^{(k)}(s_0)/k!$.
Experimentally, one usually does not have the derivatives at some point,
$s=s_0$, but a series of experimental points $F_j$ at different $s_j$, from which
one extracts the function and its derivatives  through
polynomial fits with  $S_N(s)$ at higher and higher order $N$.

But, what happens if there is a single pole at $s=s_p$ within the disk
and $F(s)$ is analytical everywhere else in $B_\delta(s_0)$?
In that case, the Taylor series does not converge any more. Nonetheless,
the needed modification is not really big. In its simplest version
with just one pole,
the de Montessus de Ballore's
theorem~\cite{Montessus,BakerPades} states that
the sequence of one-pole
Pad\'e approximants $P^N_1$ around $s_0$  converges
to $F(s)$ in any compact subset of the disk excluding the~pole~$s_p$:
\begin{eqnarray}
P^N_1(s;s_0) &=&
\sum_{k=0}^{N-1} a_k\, (s-s_0)^k\, \,
+\,\,
\Frac{a_N\, (s-s_0)^N}{1\,\,-\,\, \Frac{a_{N+1}}{a_N} \, (s-s_0)   }\, .
\end{eqnarray}
Hence, one finds that the Pad\'e pole  $x_p=s_0+\frac{a_N}{a_{N+1}}$
converges  to $s_p$ for $N\to\infty$.
Experimentally, as referred before, one is not provided with the
derivatives $F(s_0)$, $F'(s_0)$... but with the   values
 $F_j$ at different $s_j$.  We use then the rational functions
 $P^N_1 $  as fitting functions (in the way
 done before with the
polynomials). As $N$ grows $P^N_1 $ gives an estimate of the series
of derivatives $\{\, F^{(k)}(s_0)\,\}$ and, hence, of the pole position
$s_p$.

\begin{figure}[!t]
\begin{center}
  \includegraphics[width=15cm]{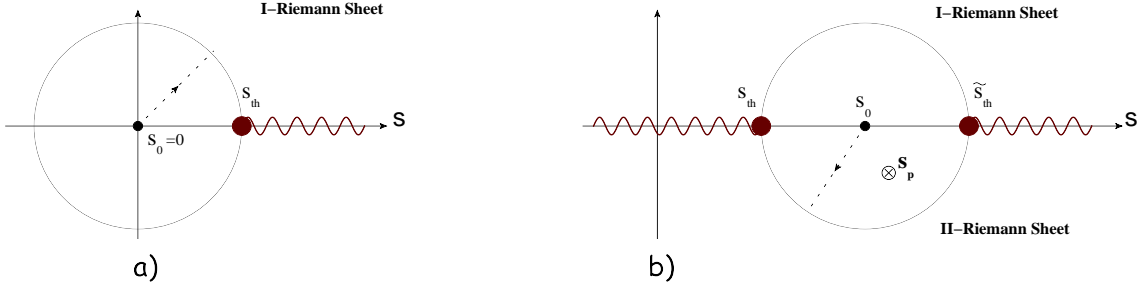}
  \caption{{\small
a) Analytical structure of the VFF bellow the first production threshold.
b) Structure and analytical extension of the $1^{st}$
Riemann sheet between the first and second production thresholds,
with a resonance pole $s_p$ in the
$2^{nd}$ Riemann sheet in the proximity of $s_0$.
  }}
\label{fig.analyt0}
\end{center}
\end{figure}

Usually, the Pad\'es have been constructed around the low-energy point
$s_0=0$ (with $s$, typically the total square momentum).
In matrix elements $F(s)$ without left-hand cut,
the amplitude is analytical from $s=-\infty$
up to the first production threshold $s_{th}$
and within the disk $B_{s_{th}}(0)$   (see Fig.~\ref{fig.analyt0}.a).
For instance,
one has $s_{th}=4m_\pi^2$ in the $\pi\pi$ vector form-factor case.
Experimentally, one may then have Euclidean data $F^{^{\rm exp}}(s)$
at $s<0$ and use them
to extract the derivatives of the VFF at $s=0$~\cite{Peris-VFF}.
Likewise, one may have Minkowskian data
$F^{^{\rm exp}}(s+i0^+)$ from $s>s_{th}$
which, strictly, cannot be used by Pad\'es centered
at $s_0=0$ due to the essential singularity at $s=s_{th}$.

Alternatively one can use in a theoretically safe way Pad\'es
centered at $s_0+i0^+$ over the brunch cut between the first and
second production thresholds, with $s_{th}<s_0<\widetilde{s}_{th}$
(see Fig.~\ref{fig.analyt0}.b).   In the $\pi\pi$--VFF  this would
correspond to the range between $s_{th}=4 m_\pi^2$ and
$\widetilde{s}_{th}=4 m_K^2$, if multipion channels are neglected.
One has then an analytical extension of our amplitude $F(s)$
in the $1^{st}$ Riemann sheet at $s+i0^+$  into
the $2^{nd}$ Riemann sheet.  Notice that the function $F(s_0+i0^+)$
and the derivatives in the $a_k$ parameters are now complex numbers.

In the case of resonant amplitudes, a single pole appears in the second
Reimann  sheet in the neighbourhood of the real $s$ axis,
which can be related to the existence of a hadronic state --resonance--
with the quantum numbers of that channel.
One can use then the de Montessus de Ballore's
theorem~\cite{Montessus,BakerPades} for the description of
the data in the maximal disk shown in Fig.~\ref{fig.analyt0}.b.
If the resonance pole lies within the disk
the $P^N_1$ Pad\'e approximants  allow its determination
in a model independent way.

\subsection{Testing the method through  various models}

We   consider a series of $\rho$--like models of the
$\pi\pi$ vector form-factor,
with a single pole in the second Riemann sheet
at $s_p=\left(0.77-\frac{i}{2} 0.15\right)^2$~GeV$^2$
and a logarithmic branch cut (starting at $s=0$ for sake of simplicity).
The considered models were
\begin{eqnarray}
\mbox{\bf Model A)}& \qquad\qquad &
F(s)\,=\,\Frac{M^2}{M^2-s +\Frac{1}{\pi}\Frac{\Gamma\, s}{M}
\ln{\Frac{-s}{M^2}}  }\, ,
\nn\\
\mbox{\bf Model B)}& \qquad\qquad &
F(s)\,=\,\Frac{M \Gamma}{M^2-s +\Frac{1}{\pi} M \Gamma
\ln{\Frac{-s}{M^2}}  }\, ,
\nn\\
\mbox{\bf Model C)}& \qquad\qquad &
F(s)\,=\,\Frac{s\ln{ \Frac{-s}{M^2} }}{
(M-i\Gamma/2)^2\,-\, s }\, ,
\end{eqnarray}
with $M$ and $\Gamma$ conveniently tuned in each case
to produce the pole at $s=s_p$.

\begin{figure}[!t]
\begin{center}
  \includegraphics[width=5cm]{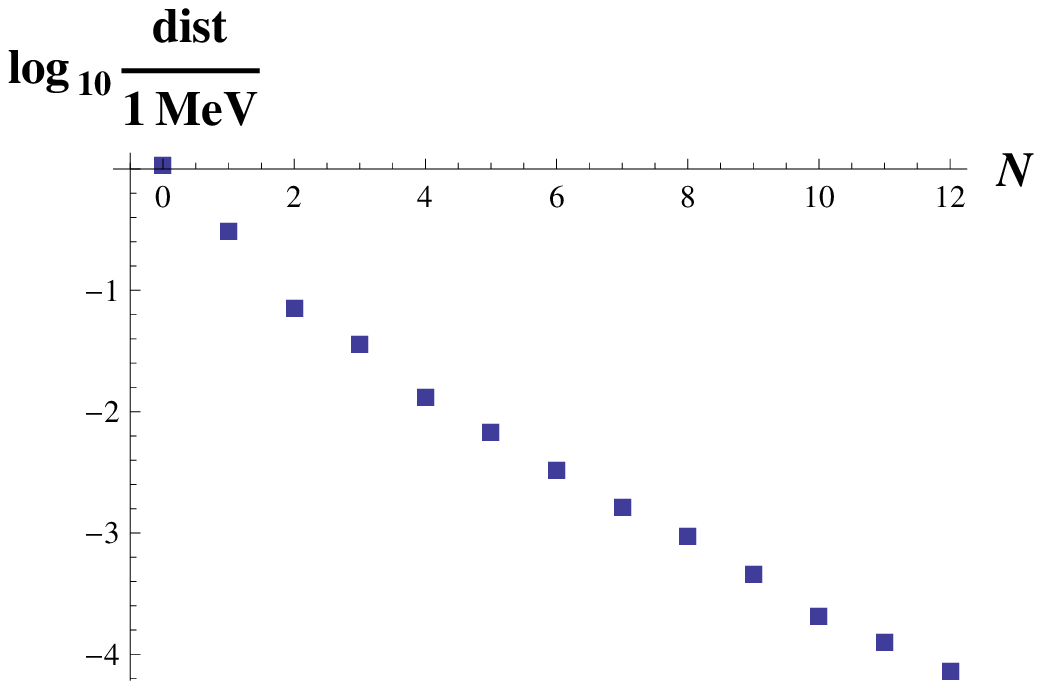}
\hspace*{0.5cm}
  \includegraphics[width=5cm]{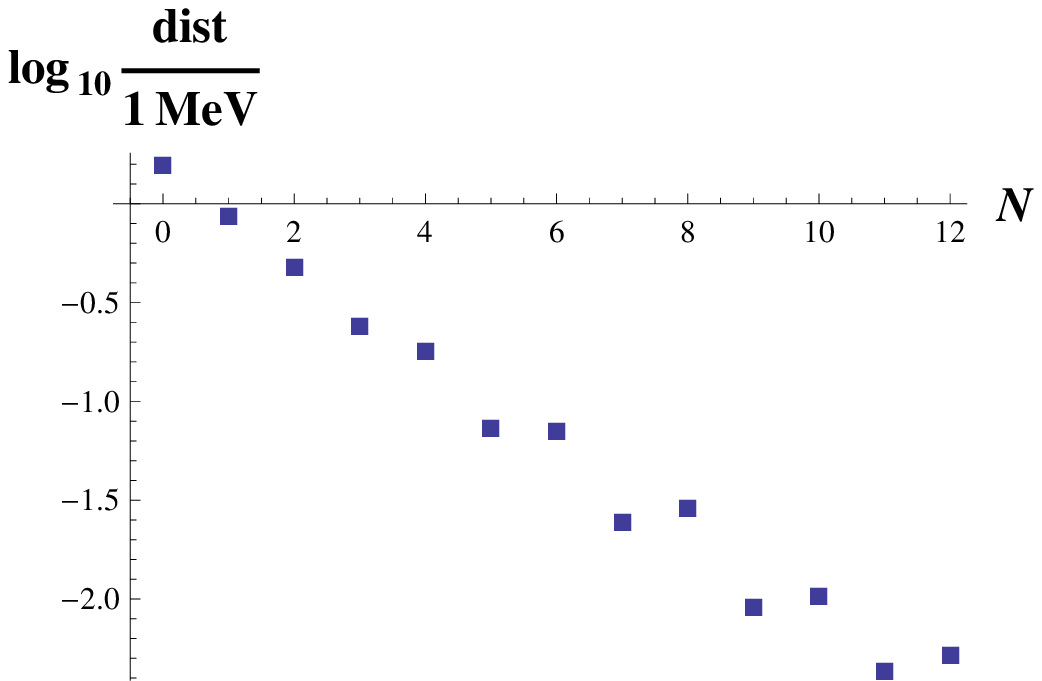}
\hspace*{0.5cm}
  \includegraphics[width=5cm]{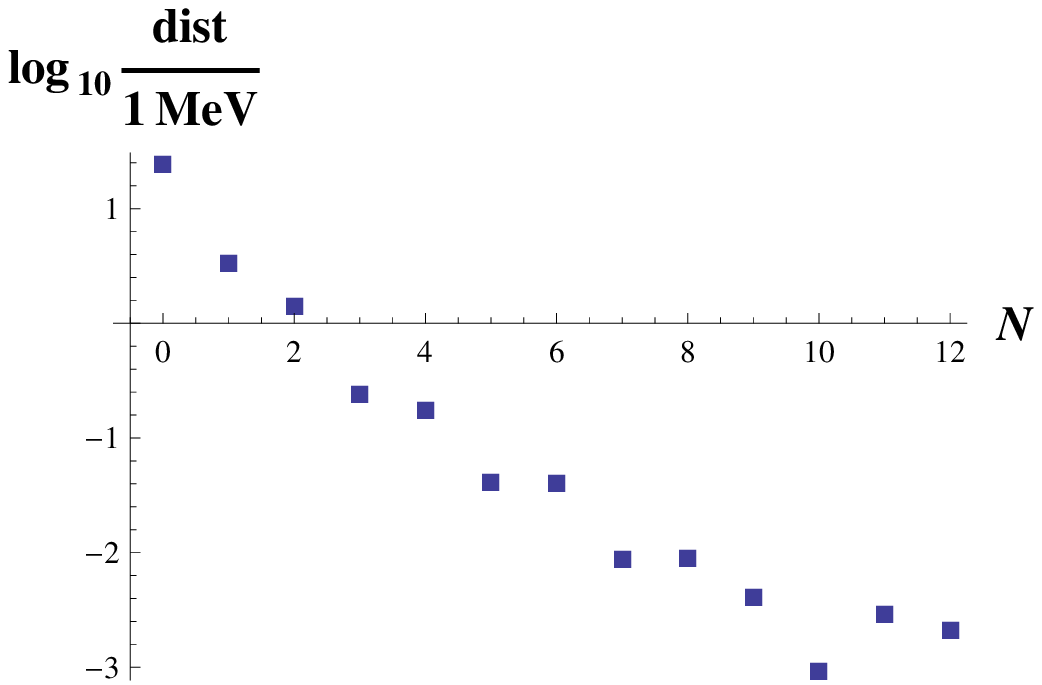}
\\
a) \hspace*{5cm} b) \hspace*{5cm} c)
\caption{{\small
a), b) and c): respectively, the
convergence of the $P^N_1$ poles to the ``physical'' value
in the models A, B and C.   The distance from the fitted
pole to the ``physical'' one is described by   $dist\equiv \left[
(M^{\rm fit}-M_p)^2+(\Gamma^{\rm fit}-\Gamma_p)^2\right]^\frac{1}{2}$.
  }}
\label{fig.Pade-poles}
\end{center}
\end{figure}

In order to simulate the physical situation we take the model
(A for instance) and generate a series of ``data'' points with zero error,
which would represent an ideal experimental situation where all the
uncertainty would be theoretical.  We fit the ``data'' for the modulus
and phase-shift of $F(s)$ and extract the optimal complex parameters
$a_k$ for each  $P^N_1$ Pad\'e.
Notice that this does not mean to fit $|F(s)|$ (or the VFF phase) with
a $P^N_1$ Pad\'e. The modulus and phase-shift of the data
are, respectively, fitted with the modulus and phase-shift of $P^N_1$.
The Pad\'e pole
$s^{^{\rm fit}}=(M^{^{\rm fit}}-i\Gamma^{^{\rm fit}}/2)^2$
is  found to converge to the ``physical''
$s_p=(M_p-i\Gamma_p/2)^2$ of the  model when $N\to\infty$.
The approaching of the complex Pad\'e pole to its limit value can be
observed in Fig.~\ref{fig.Pade-poles}.
It is clear that the rate of converge depends  on the kind of model,
being faster for model A and slower for B and C.

\subsection{Application to experimental data}

\begin{figure}[!b]
\begin{center}
  \includegraphics[width=6.5cm]{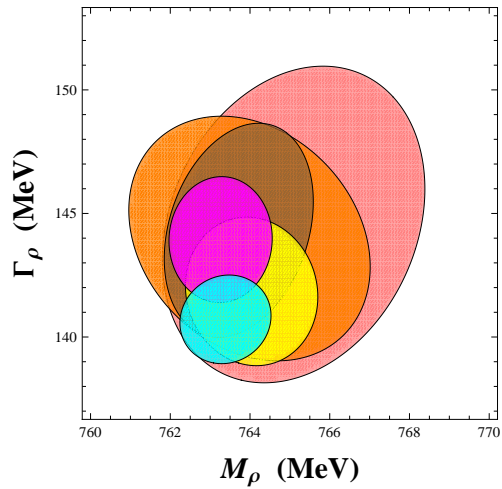}
\caption{{\small
68\% CL regions for the rho pole mass and width from
the different $P^N_1$ fits.
The smallest (cyan) ellipse provides the prediction for $N=3$ and
the following growing
orders in $N$  are given by the ovals with larger and larger size.
  }}
\label{fig.CL}
\end{center}
\end{figure}

We proceed now to analyze the final compilation of ALEPH $\pi\pi$
vector form-factor data for the squared modulus
$|F_{\pi\pi}(q^2)|^2$~\cite{ALEPH}
and the $I=J=1$ $\pi\pi$ scattering phase-shift $\delta_{\pi\pi}$,
identical to the $\pi\pi$ vector form-factor phase-shift in
the elastic  region $4m_\pi^2<q^2<4 m_K^2$
(if multipion channels are neglected).
This will be the range of application of $P^N_1$ Pad\'e analysis.
For $N\geq 3$ the fit $\chi^2$ already lies within
the 68\% confidence level (CL)  and becomes statistically acceptable.
Their corresponding 68\% CL regions for the pole mass and width predictions
are shown in Fig.~\ref{fig.CL}.
The regions from the different fits  overlap each other
in a compatible way.  The allowed ranges become larger and
larger as $N$ grows and the fit contains more and more free parameters.

At this point one needs to reach a compromise. On one hand the
experimental (fit) errors have an statistical origin and increase
as one considers higher order Pad\'es $P^N_1$,
with a larger number of parameters.  On the other,  the
systematic theoretical (Pad\'e)  error decreases as $N$  increases and the
Pad\'e converges to the actual VFF.
In the present work we have taken $N=6$ as our  best
estimate as the new parameters of Pad\'es with $N\geq 7$
turn out to be all
compatible with zero, introducing no information with respect to $P^6_1$.
Furthermore,  the different models studied before show  that in any case
the theoretical errors for mass and width
result smaller than $10^{-1}$--$10^{-2}$~MeV  for $N\geq 6$, being negligible compared to
the $\cO(1$~MeV$)$  experimental errors.
This yields the determinations
\begin{equation}
M_\rho\,=\, 763.7\pm 1.2\, \mbox{MeV}\,,
\qquad\qquad
\Gamma_\rho\,=\, 144\pm 3 \,\mbox{MeV}\, ,
\end{equation}
which is found in reasonable agreement with former
determinations obtained from more elaborated procedures
and  with similar size for the uncertainties:
\begin{eqnarray}
\mbox{[Ananthanarayan {\it et al.}~\cite{Roy} ]}
\qquad && M_p\,=\, 762.5\pm 2 \, \mbox{MeV}\,,
\qquad\qquad\quad
\Gamma_p\,=\, 142\pm 7\,\mbox{MeV}\, ,
\nn\\
\mbox{[IAM~\cite{IAM-pole-pred} ]}
\qquad && M_p\,=\, 754\pm 18\, \mbox{MeV}\,,
\qquad\qquad\quad
\Gamma_p\,=\, 148 \pm 20\,\mbox{MeV}\, ,
\nn\\
\mbox{[Zhou {\it et al.}~\cite{rho-Zheng} ]}
\qquad && M_p\,=\, 763.0\pm 0.2\, \mbox{MeV}\,,
\qquad\qquad
\Gamma_p\,=\, 139.0\pm 0.5\,\mbox{MeV}\, ,
\nn\\
\mbox{[Pich and SC~\cite{Cillero-VFF} ]}
\qquad && M_p\,=\, 764.1\pm 2.7^{+4.0}_{-2.5}\, \mbox{MeV}\,,
\qquad\quad
\Gamma_p\,=\, 148.2\pm 1.9^{+1.7}_{-5.9}\,\mbox{MeV}\, .
\end{eqnarray}

\section{Conclusions}

The Pad\'e approximants are important tools for the
analysis of QCD amplitudes. They provide alternative determinations
with competitive precision.

At low energies, we have been able to determine  the
first derivatives of the $\pi\pi$--VFF at the origin, this is,
its quadratic charge radius $\bra r^2\ket_V^\pi$  ($=6 a_1$)
and the curvature $c_V^\pi$ ($= a_2$).
For this, we used   Euclidian data, i.e., from $q^2<0$.

Likewise, we were able to employ the Pad\'e approximants
to extract information
about the  hadronic resonance poles (the $\rho(770)$ mass and width).
The experimental Minkowskian VFF and $\pi\pi$--scattering data
were analyzed  by means of $P^N_1$ Pad\'es centered
between the first and second production thresholds.
We obtained the determinations
$M_\rho=763.7\pm 1.2$~MeV and
$\Gamma_\rho=144\pm 3$~MeV,  with
a competitive precision compared to other more elaborated and complex
methods,  in spite of the simplicity of the proposed procedure.

This study shows the Pad\'e approximants, once again, as a useful tool
for the investigation of QCD phenomenology.
They provide alternative determinations and,
in spite of their simplicity,
they have been proven as an efficient and systematic instrument.
Nevertheless,
when no theorem supports the convergence of the Pad\'e sequence,
the extracted parameters may have little to do with the
physical ones, as we saw in our LSM study of ``Pad\'e''--unitarizations
in~Sec.~3.

\vspace*{1cm}





\end{document}